\documentclass[aps,prc,twocolumn,final,showpacs,twoside]{revtex4-1}
\usepackage{amsmath,amsfonts,amssymb,bm}
\usepackage{mathptmx}
 \DeclareMathAlphabet{\mathbi}{OT1}{ptm}{bx}{it}
 \DeclareSymbolFont{symbols}{OMS}{cmsy}{m}{n}
\usepackage[dvips]{graphicx}

\newcommand{\etal}{\textit{et al.}}
\newcommand{\bp}{\mathbf{p}}
\newcommand{\br}{\mathbf{r}}
\newcommand{\rA}{\mathrm{A}}
\newcommand{\rB}{\mathrm{B}}
\newcommand{\rE}{\mathrm{E}}
\newcommand{\rF}{\mathrm{F}}
\newcommand{\rTd}{\bm{T}_d}
\newcommand{\cM}{\mathcal{M}}
\newcommand{\cE}{\varepsilon}

\newcommand{\pp}[2]{\frac{\partial #1}{\partial #2}}
\newcommand{\Matrix}[1]{\begin{pmatrix}#1\end{pmatrix}}
\newcommand{\btetra}{\beta_{\rm td}}

\begin{document}
\title{Semiclassical origin of anomalous shell effect for 
tetrahedral deformation \\
in radial power-law potential model}
\author{Ken-ichiro Arita and Yasunori Mukumoto}
\affiliation{Department of Physics, 
Nagoya Institute of Technology, Nagoya 466-8555, Japan}

\received{March 17, 2014}

\begin{abstract}
Shell structures in single-particle energy spectra are investigated
against regular tetrahedral type deformation using a radial power-law
potential model.  Employing a natural way of shape parametrization
which interpolates sphere and regular tetrahedron, we find prominent
shell effects for rather large tetrahedral deformations, which bring
about shell energies much larger than the cases of spherical and
quadrupole type shapes.  We discuss the semiclassical origin of these
anomalous shell structures using periodic orbit theory.
\end{abstract}


\maketitle

\section{Introduction}

Recent progress in experimental facilities has opened the new
frontiers of unstable nuclei, where the combinations of neutron and
proton numbers are considerably distant from the $\beta$ stability
line.  Various kinds of exotic shapes are expected in several regions
of deformed doubly magic nuclei as well as $N=Z$ regions ($N$ and $Z$
being neutron and proton numbers, respectively), where neutron and
proton shell effects play cooperative roles.  Tetrahedral deformed
states are one of the candidates for such exotic
states \cite{Takami,Dudek,Tagami}.  For such states, single-particle
levels acquire degeneracies due to the high point-group symmetry of
the Hamiltonian, and single-particle spectra show extra shell effects
in comparison to the other types of deformations.  Hamamoto
\etal compared the four types of octupole deformations $Y_{3m}$ ($m=0$,
1, 2, and 3) in the modified oscillator model (Nilsson model without
spin-orbit term), and found a stronger shell effect in $Y_{32}$ shape,
which possesses tetrahedral symmetry, in comparison to the other three
\cite{Hamamoto}.  They emphasized the occurrence of global level
bunchings as well as the level degeneracies due to the point-group
symmetry.  In the analysis of electronic shell structures in sodium
clusters with jellium model, notable shell effects were found for
tetrahedral deformed states \cite{Reimann}.  Interestingly, the
resulting tetrahedral magic numbers, $N=2$, 8, 20, 40, 70, 112,
$\ldots$, are exactly the same as those for the spherical harmonic
oscillator (HO) model.  Dudek \etal investigated tetrahedral shell
structures using the realistic nuclear mean-field model (including
spin-orbit and Coulomb terms) with a more elaborate way of shape
parametrization by adopting several combinations of spherical
harmonics which are symmetric with respect to any transformation of a
tetrahedral group \cite{Dudek}.  They also found strong level bunchings
for finite tetrahedral deformation.  The deformed magic numbers are
shifted from the HO values, which might be mainly due to the
spin-orbit coupling.  These level bunchings may suggest a restoration
of dynamical symmetry by tetrahedral deformation, which are broken in
a spherical potential with a sharp surface.

In Sec.~\ref{sec:quantum}, we show the results of tetrahedral deformed
shell structures using the radial power-law potential model, which we
proposed as an approximation to the Woods-Saxon
model \cite{Arita2004,Arita2012}.  One will see remarkable shell
structures emerging in transition from spherical to regular
tetrahedral shapes.  Semiclassical periodic orbit theory is applied to
clarify the mechanism which brings about the above significant
enhancement of shell effects.  Semiclassical theory of shell
structures are described in Sec.~\ref{sec:theory}.  Special attention
will be paid to the significance of periodic-orbit bifurcations.
Trace formula for the radial power-law potential model is derived in
Sec.~\ref{sec:scaling}.  Properties of classical periodic orbits in
the tetrahedral deformed potential are studied in
Sec.~\ref{sec:orbit}, and the quantum-classical correspondences are
investigated in Sec.~\ref{sec:semiclassical}.
Section~\ref{sec:summary} is devoted for summary and discussions.

\section{Quantum shell structures in transition from
sphere to tetrahedron}
\label{sec:quantum}

If one parametrizes the tetrahedral type shapes only with a $Y_{32}$
term, the shape becomes quite strange for large deformation, and
classical dynamics turns strongly chaotic due to the negative
curvature of the potential surface.  Usually, one may not expect
deformed shell structures to develop in such chaotic systems.  In the
present analysis, we make use of a simple and more natural way of
shape parametrization which smoothly interpolates spherical and regular
tetrahedral shapes with one parameter.  We adopt the power-law type
radial dependence of the potential, $V\propto r^\alpha$, which nicely
approximates the Woods-Saxon type potentials inside the surface for
systems with a wide range of constituent particle
numbers \cite{Arita2004,Arita2012}.  This choice of the radial
dependence is useful in both quantum and semiclassical analyses
because of the scaling properties of the
system \cite{Arita2004,Arita2012,Magner2013}.

\begin{figure}[tb]
\includegraphics[width=\linewidth]{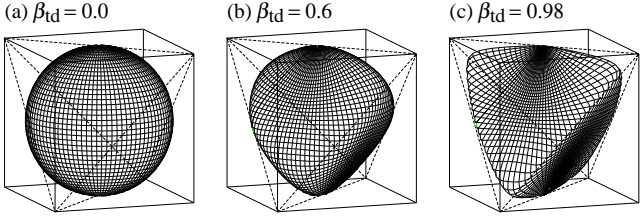}
\caption{\label{fig:vshape}
Shapes of equipotential surface given by Eq.~(\ref{eq:shape_fun})
with $\btetra=0$ (a), $0.6$ (b), and $0.98$ (c).}
\end{figure}

The radial power-law potential model is described by the Hamiltonian
\begin{equation}
H=\frac{p^2}{2m}+U_0\left(\frac{r}{R_0f(\theta,\varphi;\btetra)}\right)^\alpha.
\label{eq:hamiltonian}
\end{equation}
Here, $m$ is the mass of the constituent particles, $U_0$ and $R_0$
are the constants having dimensions of energy and length,
respectively, which will be used as the units of those quantities.
[$U_0$ is not an independent constant and we will fix it later: see
Eq.~(\ref{eq:vars}) below.]  $f(\theta,\varphi;\btetra)$ is the
tetrahedral shape function with deformation parameter $\btetra$, given
implicitly by the least positive root of the fourth order equation
\begin{equation}
f^2+\frac{\btetra}{2}\left(1+u_3(\theta,\varphi)f^3
 -u_4(\theta,\varphi)f^4\right)=1 \label{eq:shape_fun}
\end{equation}
with
\begin{gather}
u_3(\theta,\varphi)=\frac{4}{15}P_{32}(\cos\theta)\sin 2\varphi, \\
u_4(\theta,\varphi)=\frac15+\frac45P_4(\cos\theta)
 +\frac{1}{210}P_{44}(\cos\theta)\cos4\varphi\,.
\end{gather}
$u_3$ and $u_4$ are symmetric under any transformation of the
tetrahedral group $\rTd$.  In Eq.~(\ref{eq:shape_fun}),
$\btetra=0$ corresponds to spherical shape ($f=1$), and $\btetra=1$
corresponds to a regular tetrahedral shape (see Appendix~\ref{app:shape}
for the detailed derivation).  Thus, one can smoothly change the shape
of the potential from a sphere to regular tetrahedron, keeping the
tetrahedral symmetry, by varying the single parameter $\btetra$ from 0
to 1.  In order to satisfy the volume conservation condition, $R_0$
should be determined as a function of $\btetra$ by
\begin{equation}
R_0(\btetra)=R_0(0)\left[\frac{1}{4\pi}\int d\varOmega
 f^3(\varOmega;\btetra)\right]^{-1/3}.
\end{equation}
Figure~\ref{fig:vshape} shows the equipotential surface for
several values of $\btetra$.
In our shape parametrization, the surface is convex everywhere for any
value of $\btetra$, and the classical orbits are less chaotic compared
with the case of pure $Y_{32}$ shapes
\begin{equation}
f(t_3)=1+t_3(Y_{32}+Y_{3-2}). \label{eq:y32}
\end{equation}
Actually, a pure $Y_{32}$ equipotential surface shows concavity at
$t_3=0.3$, a value where deformed shell effects are most enhanced,
and it turns more eccentric for larger $t_3$.

The quantum spectra are calculated by diagonalizing the Hamiltonian with
harmonic oscillator basis.  After suitable scale transformations of
variables, the Schr\"{o}dinger equation for single-particle energy $e$ is
transformed into the following dimensionless form:
\begin{equation}
\left[-\frac12\bar{\bm{\nabla}}^2+\left(\frac{\bar{r}}{f}\right)^\alpha\right]
\psi(\bar{\br})=\bar{e}\psi(\bar{\br}),
\end{equation}
with
\begin{equation}
\bar{\br}=\frac{\br}{R_0}, \quad \bar{e}=\frac{e}{U_0}, \quad
U_0=\frac{\hbar^2}{mR_0^2}. \label{eq:vars}
\end{equation}
$\bar{\bm{\nabla}}$ represents the derivative with respect to the
dimensionless coordinate $\bar{\br}$.  In the following part, we shall
omit the bars on those variables and operators for simplicity.  The
diagonalization can be carried out in each block of irreducible
representation (irrep) of $\rTd$ \cite{LLQMText,GroupText}.  The group
$\rTd$ has five irreps including three-dimensional representations.
The $f$-dimensional irrep leads to the spectrum with $f$-fold
degeneracy.  In the diagonalization process, we can further decompose the
bases into those of no degeneracy.  A detailed procedure of the
decomposition of bases making use of the $\rTd$ symmetry is presented
in Appendix~\ref{app:group}.

\begin{figure}[tb]
\begin{center}
\includegraphics[width=.85\linewidth]{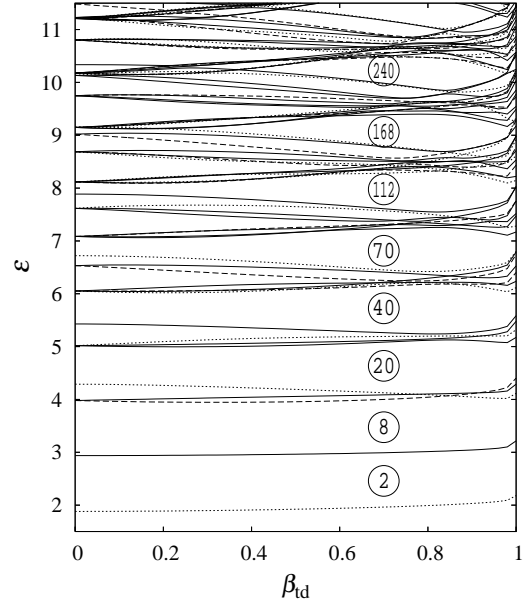}
\end{center}
\caption{\label{fig:spectrum}
Single-particle level diagram for radial parameter $\alpha=6.0$.
Scaled energy eigenvalues $\cE_i=e_i^{1/2+1/\alpha}$ [see
Eq.~(\ref{eq:scaledvar})] are plotted as
functions of tetrahedral deformation parameter $\btetra$.  Dotted, dashed
and solid curves represent levels belonging to irreps $\rA_{1,2}$,
$\rE$ and $\rF_{1,2}$, respectively.  Circled numbers indicate magic
numbers for the closed-shell configuration, taking the spin
degeneracy factor into account.}
\end{figure}

For $\alpha$ close to 2, outstanding shell effects at the spherical
shape (isotropic HO) are monotonically reduced by increasing
tetrahedral deformation up to $\btetra=1$.  For larger $\alpha$, namely
for potentials with sharper surfaces, spherical shell structures
become moderate.  However, we found the emergence of remarkable shell
effects at rather large tetrahedral deformation.
Figure~\ref{fig:spectrum} shows a single-particle level diagram for the
case of radial parameter $\alpha=6$, where quantum energy spectra are
plotted against tetrahedral parameter $\btetra$.  One sees strong
bunchings of levels and the appearance of large energy gaps around
$\btetra\approx 0.7$.  The effect looks much more remarkable than those
found in Refs.~\cite{Dudek,Hamamoto}.
In order to compare the above tetrahedral deformation
with the popular shape parametrization, let us expand our shape
function $f(\btetra)$ by the spherical harmonics as in
Eq.~(\ref{eq:y32}).  One obtains the parameter $t_3$ by
\begin{equation}
t_3=\int \tfrac12(Y_{32}+Y_{3-2})f(\btetra)d\varOmega. \label{eq:t3}
\end{equation}
For $\btetra=0.7$, we have $t_3=0.27$, which is close to the value
for which tetrahedral shell effects are considerably enhanced in
Ref.~\cite{Dudek}.

One will also note that the deformed magic numbers are exactly the same as
those for the spherical harmonic oscillator, namely, $N=2$, 8, 20, 40,
$\ldots$, as obtained in Ref.~\cite{Reimann}.  These numbers are not
changed by varying the value of parameter $\alpha$ in a range
$\alpha\gtrsim 4$, while the deformation parameter $\btetra$ at which
the shell effect becomes most remarkable is shifted to a larger value as
$\alpha$ increases.

The above situation also reminds us of the deformed shell structure in
a two-dimensional billiard system where strong level bunchings are found
in transition from circular to equilateral triangular
shape \cite{TriCirc}.  The origin of this anomalous shell effect was
understood as the bifurcation enhancement effect of short classical
periodic orbits.  This strongly encourages us to investigate the
periodic orbits in our tetrahedral model and make a semiclassical
analysis using periodic orbit theory.

\section{Semiclassical theory of shell structures\\
--- The role of the periodic orbit bifurcations}
\label{sec:theory}

In semiclassical trace formula, the quantum energy level density $g(e)$
is expressed as the sum over contribution of classical periodic
orbits \cite{Gutzwiller}
\begin{gather}
g(e)=g_0(e)+\sum_k\sum_{n=1}^\infty  g_{nk}(e),
\label{eq:trace_formula} \\
g_{nk}(e)=A_{nk}(e)\cos\left[\frac{n}{\hbar}S_k(e)
 -\frac{\pi}{2}\sigma_{nk}\right].
\label{eq:trace_formula2}
\end{gather}
The sum in the right-hand side of Eq.~(\ref{eq:trace_formula}) is
taken over all primitive classical periodic orbits $k$ and their $n$-th
repetitions.  $S_k=\oint_k\bp\cdot d\br$ is the action integral along
the orbit $k$, $\sigma_{nk}$ is the Maslov index for $n$-th repetition
of the orbit $k$ which is an integer related with the geometric property
of the orbit.  For an isolated orbit, the amplitude factor $A_{nk}(e)$ is
given by the Gutzwiller formula
\begin{equation}
A_{nk}(e)=\frac{N_kT_k}{\pi\hbar\sqrt{|\det(I-M_{nk})|}},
\label{eq:amp_gutz}
\end{equation}
where $T_k$ is the period of the primitive orbit $k$, and
$M_{nk}={M_k}^n$ is the monodromy matrix explained below.  $N_k$ is
the multiplicity due to the discrete symmetry of the system, namely,
there are $N_k$ identical copies of the orbit $k$ generated by the
symmetry transformations.  $g_0(e)$ represents the average level density
given by the Thomas-Fermi (TF) approximation \cite{BBText},
\begin{equation}
g_{\rm TF}(e)=\frac{1}{(2\pi\hbar)^3}\int d\bp d\br
 \delta(e-H(\bp,\br)).
\end{equation}

The energy of the $N$ particle system, $E(N)$, is decomposed into a
smooth part and an oscillating part as \cite{Strutinsky}
\begin{equation}
E(N)=\tilde{E}(N)+\delta E(N),
\end{equation}
and the oscillating part, referred to as shell correction energy, is
related with the oscillating part of the level density $\delta g(e)$
as \cite{StrutMagner,BBText}
\begin{equation}
\delta E(N)=\int_{-\infty}^{e_F}(e-e_F)\delta g(e)de,
\end{equation}
where $e_F$ is the Fermi energy satisfying
\begin{equation}
\int_{-\infty}^{e_F}g(e)de=N.
\end{equation}
Inserting the semiclassical expression
$\delta g(e)=\sum_{nk}g_{nk}(e)$, one obtains \cite{StrutMagner,BBText}
\begin{equation}
\delta E(N)=\sum_{nk}
\left(\frac{\hbar}{nT_k(e_F)}\right)^2g_{nk}(e_F).
\label{eq:eshell}
\end{equation}
As seen from the above equation, longer periodic orbits have less
contribution to the shell correction energy due to the factor
$(nT_k)^{-2}$.  Thus, the shell correction energy is dominated by the
gross shell structure related with short periodic orbits.

In a three-dimensional system, calculations of classical periodic orbits
require the search in a four-dimensional Poincar\'{e} section.  The energy
$e$ defines a five-dimensional hypersurface in six-dimensional phase space.
Let us define an appropriate four-dimensional hypersurface $\Sigma$ in
this five-dimensional energy surface.  In the calculations below, we
consider the hypersurface
$\Sigma=\{Z\equiv(x,y,p_x,p_y)|z=0,\dot{z}>0\}$, $p_z$ being
determined by the energy condition $H(\bp,\br)=e$.  
Classical dynamics defines the mapping
$\cM:\Sigma\mapsto\Sigma$ (Poincar\'{e} map)
\begin{equation}
Z'=\cM(Z), \quad Z, Z'\in \Sigma,
\end{equation}
where a classical orbit started at $Z$ on $\Sigma$ intersects
$\Sigma$ again later at $Z'$.  The periodic orbits $Z_k$ of period $p$
are fixed points of the $p$\,th power of the Poincar\'{e} map,
\begin{equation}
Z_k=\cM^p(Z_k),
\end{equation}
and the monodromy matrix $M_k$ is the linearized Poincar\'{e} map around the
periodic orbit
\begin{gather}
\delta Z_k'=\cM^p(Z_k+\delta Z_k)-Z_k
\simeq \left.\pp{\cM^p}{Z}\right|_{Z_k}\delta Z_k
\equiv M_k\delta Z_k.
\end{gather}
For $n$-th repetitions, one has
\begin{equation}
\delta Z'_{nk}=M_k\delta Z'_{(n-1)k}=\cdots={M_k}^n \delta Z_k.
\end{equation}
Periodic orbit bifurcations occur when the monodromy matrix has unit
eigenvalue, namely at $\det(I-{M_k}^n)=0$.  At the bifurcation
deformation, the periodic orbit forms a local family of quasiperiodic
orbits in the direction of the eigenvector belonging to the unit
eigenvalue of the monodromy matrix.  They make a coherent contribution
to the level density and sometimes lead to a significant enhancement of
the amplitude $A_{nk}$.  This semiclassical mechanism
nicely explains the enhancement of the deformed shell structure in many
studies \cite{Arita95,Sugita98,Arita98,Magner99,Magner2002}.  It also
suggests a restoration of local dynamical symmetry, namely, the
symmetry transformation generates a family of periodic orbits having
the same periods and stabilities like, for instance, a degenerate family
of orbits in rotationally symmetric potentials.

\begin{figure}
\includegraphics[width=\linewidth]{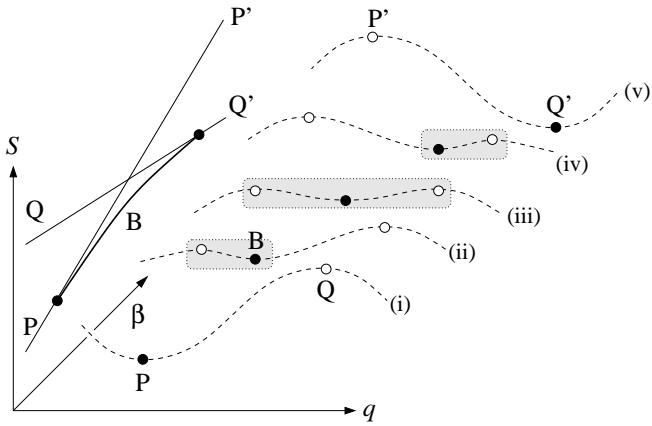}
\caption{\label{fig:bridgebif}
Illustration of bridge orbit bifurcation scenario where orbit B
bridges the two periodic orbits P and Q.  Dashed curves represent the
action integral $S(q)$ along the closed orbit for several values of
deformation $\beta$.  Its stationary points indicated by solid and
open dots correspond to the stable and unstable periodic orbits,
respectively.  In the left-hand side, action integrals along those
periodic orbits are shown by solid curves as functions of deformation
$\beta$.}
\end{figure}

A substantial enhancement of the shell effect is expected in accordance
with the so-called bridge orbit bifurcations.  Typically, bridge
orbits emerge between two quite different periodic orbits when their
periods cross in the deformation-action ($\beta-S$) plane \cite{Bridge}.
For instance, consider the two-dimensional anisotropic harmonic
oscillator (HO) perturbed by a nonlinear term.  In pure (unperturbed)
HO system, there are two isolated diametric orbits along principal
axes.  When the two oscillator frequencies encounter in a rational
ratio, all the classical orbits of Lissajous figures become periodic
with same periods and stabilities, namely, they form a doubly degenerate
family.  This is related with the dynamical symmetry \cite{rationalHO}.
If a perturbation is imposed and the above dynamical symmetry is
slightly broken, the degenerate family will be replaced by the
bridge orbit(s), which emerge from one diameter at smaller
deformation and then submerge into the other diameter at larger
deformation.  Since the loci or two diameters in phase space are quite
distant from each other, bridge orbits will travel a long way in the
phase space from one to the other during the bifurcation scenario.
This is in contrast to usual bifurcations where all the participating
periodic orbits reside in the vicinity of the central periodic orbit.
Figure~\ref{fig:bridgebif} illustrates an example of the bridge orbit
bifurcation scenario.  Action integral $S(q)$ along the closed orbit
which starts $q$ and returns to the same point $q$ is shown as a
function of $q$ and deformation $\beta$.  The stationary points
indicated by solid and open circles correspond to periodic orbits.
With increasing $\beta$, bridge orbit bifurcation occurs as follows:
\begin{enumerate}\itemsep=0pt\def\labelenumi{(\roman{enumi})}
\item There are two periodic orbits P (stable) and Q (unstable).
\item The stable orbit P bifurcates and a new orbit B emerges.  P turns
unstable (P') afterwards.  A local family of quasiperiodic orbits
 (indicated by the shaded range) are formed around orbits P and B.
\item Orbits P and Q cross in the $\btetra-S$ plane.  The
quasiperiodic orbit family extends from P to Q.
\item Orbit B approaches the orbit Q.
\item Orbit B submerges into the orbit Q, and Q becomes stable (Q')
afterwards.
\end{enumerate}
Thus, due to the coherent contribution of the quasiperiodic orbit family,
the bridge orbit is expected to make a significant contribution to the level
density and leads to an enhancement of the shell effect.  Since the family
formed around the bridge orbit is extended between two orbits P and Q
which are distant from each other in the phase space, this bifurcation
is related with the restoration of a somewhat global dynamical symmetry
in comparison to the simple bifurcations.  Note that the bridge orbit does
not appear for every crossing of two orbits in $(\beta-S)$ plane.
The appearance of bridge orbits might be an indication of the
existence of global approximate dynamical symmetry.

\section{Trace formulas for the radial power-law potential model}
\label{sec:scaling}

Using the scaling invariance of our radial power-law potential model,
the semiclassical analysis becomes quite simple.  Since the
Hamiltonian (\ref{eq:hamiltonian}) has the scaling property
\begin{equation}
H(c^{1/2}\bp,c^{1/\alpha}\br)=cH(\bp,\br),
\end{equation}
the equations of motion are invariant under the scale transformation
\begin{gather}
\bp\to c^{1/2}\bp,\quad
\br\to c^{1/\alpha}\br,\quad
t\to c^{1/\alpha-1/2}t,\nonumber \\
\text{with}\quad
e\to ce.
\end{gather}
Therefore, classical dynamics does not depend on energy $e$, and
classical trajectories at any energy $e$ can be obtained from those
at the certain reference energy $e_0$ by the scale transformation
\begin{gather}
(\bp(t),\br(t))_e=(c^{1/2}\bp(t'),c^{1/\alpha}\br(t'))_{e_0}
\nonumber\\ \text{with}\quad
c=\frac{e}{e_0}, \quad t=c^{1/2-1/\alpha}t'.
\end{gather}
Thus we have the same set of periodic orbits regardless of energy.
The action integral along the periodic orbit $k$ is written as
\begin{equation}
S_k(e)=\oint_{k(e)}\bp\cdot d\br
=\left(\frac{e}{e_0}\right)^{1/2+1/\alpha}\oint_{k(e_0)}\bp\cdot d\br
=\cE\hbar\tau_k.
\end{equation}
In the last equation, we define dimensionless \textit{scaled
energy} $\cE$ and \textit{scaled period} $\tau_k$;
\begin{equation}
\cE\equiv(e/e_0)^{1/2+1/\alpha}, \quad
\tau_k\equiv S_k(e_0)/\hbar, \label{eq:scaledvar}
\end{equation}
which are proportional to the
original energy $e$ and period $T_k$, respectively, in the HO limit
$\alpha\to 2$.  Then, the semiclassical trace formula for
the scaled-energy level density becomes
\begin{equation}
g(\cE)=g(e)\frac{de}{d\cE}
\simeq g_0(\cE)+\sum_{nk}a_{nk}(\cE)\cos(n\cE\tau_k-\tfrac{\pi}{2}
\sigma_{nk}),
\label{eq:scaled_dos}
\end{equation}
where amplitude factor $a_{nk}(\cE)$ is related with the
original one $A_{nk}(e)$ by
\begin{equation}
a_{nk}(\cE)=\frac{de}{d\cE}A_{nk}(e(\cE)),\quad
e(\cE)=e_0\cE^{\frac{2\alpha}{\alpha+2}}.
\end{equation}
Using the Gutzwiller formula (\ref{eq:amp_gutz}), the amplitude
$a_{nk}$ becomes
\begin{equation}
a_{nk}=\frac{N_k\tau_k}{\pi\sqrt{|\det(I-M_{nk})|}}
\end{equation}
which is independent of energy.
The shell correction energy (\ref{eq:eshell}) becomes
\begin{align}
\delta E(N)&\simeq\left.\frac{de}{d\cE}\right|_{\cE_F}
 \int^{\cE_F}(\cE-\cE_F)\delta g(\cE)d\cE \nonumber \\
 &=\left.\frac{de}{d\cE}\right|_{\cE_F}
 \sum_{nk}\frac{1}{(n\tau_k)^2}a_{nk}
 \cos(n\tau_k\cE_F-\tfrac{\pi}{2}\sigma_{nk})
\end{align}
with
\begin{align}
N&=\int^{\cE_F}\left(g_0(\cE)+\sum_{nk}a_{nk}
 \cos(n\tau_k\cE-\tfrac{\pi}{2}\sigma_{nk})\right)d\cE \nonumber \\
 &=N_0(\cE_F)+\sum_{nk}\frac{1}{n\tau_k}a_{nk}
 \sin(n\tau_k\cE_F-\tfrac{\pi}{2}\sigma_{nk}).
\end{align}
$N_0(\cE)$ is the average number of levels below the scaled
energy $\cE$.  In the Thomas-Fermi approximation, $g_0$ and $N_0$ are
given by
\begin{gather}
g_{\rm TF}(\cE)=c_0\cE^2, \quad
c_0=\frac{2\sqrt{2}}{\pi}\rB\left(1+\frac{3}{\alpha},\frac32\right),\\
N_{\rm TF}(\cE)
=\int_0^{\cE}g_{\rm TF}(\cE')d\cE'=\frac{c_0}{3}\cE_F^3,
\end{gather}
where $\rB(s,t)$ is the Euler's beta function \cite{Arita2012}.  Note
that $g_{\rm TF}(\cE)$, and hence $N_{\rm TF}(\cE)$, is independent of
deformation under the volume-conservation condition.  Since $\delta E$ and
$N$ are both uniquely determined as functions of $\cE_F$, one can
obtain $\delta E$ of $N$ by using $\cE_F$ as the parameter.

\section{Classical periodic orbits}
\label{sec:orbit}

\begin{figure}[tb]
\includegraphics[width=\linewidth]{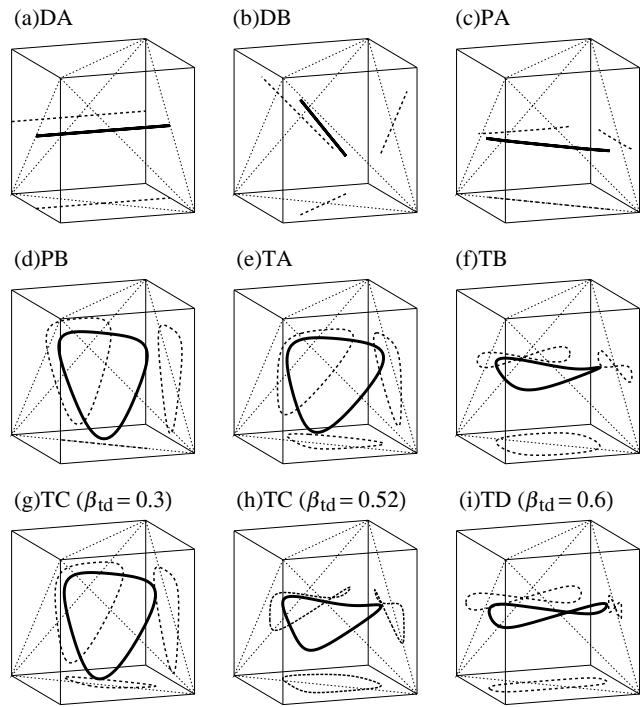}
\caption{\label{fig:orbits}
Some short classical periodic orbits calculated for $\alpha=6.0$.
(a)--(f) are the six shortest orbits which exist at small $\btetra$,
drawn here for $\btetra=0.3$.  (g)--(i) are some bifurcated orbits
for given $\btetra$.  In each panel, the thick solid curve represents the
orbit and their projections onto the three faces of the outer cube are
drawn with thick dashed curves.  The tetrahedron drawn with dotted
lines indicates the symmetry of the potential.}
\end{figure}

In order to estimate semiclassical level densities and shell
correction energies by means of the trace formula, let us study the
properties of short periodic orbits in a tetrahedral potential.  In the
spherical limit $\btetra=0$ with $\alpha>2$, there are two families of
shortest orbits, diametric and circular ones, each of which are doubly
degenerated.  With increasing $\alpha$, triply degenerated regular
polygonal-type families of orbits bifurcate from the circular one.
The first such family is the triangular type one which emerges at
$\alpha=7$.  For $\alpha=6$, the shortest orbits are only diameters
and circles.  Imposing the tetrahedral deformation, each of two
families bifurcate into three kinds of isolated orbits.  The diametric
family bifurcates into diameter `DA' along $S_4$ (four fold rotatory
reflection) axis, diameter `DB' along $C_3$ (three fold rotation) axis,
and librational `PA' in the symmetry planes.  These orbits have several
identical copies generated by the symmetry transformations.  The
multiplicities of DA, DB, and PA orbits are 3, 4, and 6, respectively.
The circular family bifurcates into isosceles triangular type `PB' in
the symmetry plane, three-dimensional `TA' having $C_3$ and $\sigma_d$
(reflection) symmetries, and three-dimensional `TB' having $S_4$ and
$\sigma_d$ symmetries.  PB, TA and TB orbits have multiplicities 12, 8
and 6, respectively.  Those six kinds of orbits for $\alpha=6$ and
$\btetra=0.3$ are illustrated in Fig.~\ref{fig:orbits}(a)--(f).
In each panel, we also show the
projections onto three faces of the outer cube and the
tetrahedron expressing the symmetry of the potential for ease of
understanding its geometry.

\begin{figure}[tb]
\includegraphics[width=\linewidth]{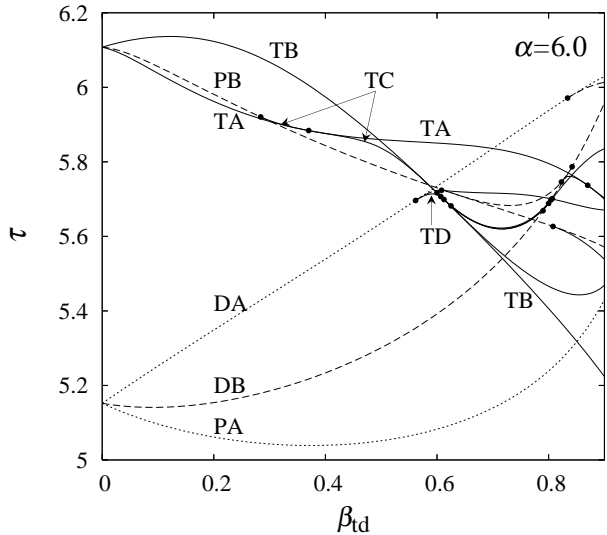}
\caption{\label{fig:tau}
Scaled periods of the shortest periodic orbits for $\alpha=6$
as functions of the tetrahedral deformation parameter $\btetra$.
Solid, dashed, and dotted curves represent three-dimensional,
planar, and diametric orbits, respectively.
Solid circles indicate bifurcation points.}
\end{figure}

In Fig.~\ref{fig:tau}, scaled periods $\tau_k$ of these shortest
periodic orbits are plotted as functions of tetrahedral deformation
parameter $\btetra$.  With increasing tetrahedral deformation, periods
and stabilities of the orbits change and some of those orbits undergo
bifurcations.  The orbit PB undergoes bifurcation at $\btetra=0.283$
and a three-dimensional orbit TC emerges from it.  In this new orbit,
all the symmetries are broken and its multiplicity is 24.  With
increasing $\btetra$, it undergoes ``touch-and-go'' (nongeneric
period-tripling) bifurcation \cite{Ozorio} with the orbit TA at
$\btetra=0.369$ and finally submerges into the orbit TB at
$\btetra=0.600$.  Namely, the orbit TC makes bridges between PB and
TA, then between TA and TB.  The orbit DA undergoes bifurcation at
$\btetra=0.562$ and three-dimensional orbit TD emerges from it.  With
increasing $\btetra$, TD submerges into the orbit TB at
$\btetra=0.607$, namely, it makes a bridge between DA and TB.  One
should note that intensive bifurcations take place around
$\btetra\sim0.6$ and many of them form bridges between crossing
periodic orbits in the region $\btetra=0.6\sim 0.8$.
As discussed in the last part of Sec.~\ref{sec:theory}, the situation
in our model, where periods of several periodic orbits come close to
each other around $\btetra=0.6$ and exchange bridge orbits between
them, may indicate a global dynamical symmetry approximately restored
in this deformation region.

\section{Semiclassical analysis of tetrahedral shell structures}
\label{sec:semiclassical}

Knowing the classical periodic orbits in the previous section, let us
proceed on to a semiclassical analysis of the deformed shell structures.
In evaluating trace formula, periods, monodromy matrices and Maslov
indices are required.  Periods and monodromy matrices are
automatically obtained in the process of calculating periodic orbits
by the monodromy method \cite{Baranger,Provost}.  In order to calculate
Maslov indices in a three-dimensional system, we devise a useful
technique which is presented in Appendix~\ref{app:maslov}.

\begin{figure}[bt]
\includegraphics[width=\linewidth]{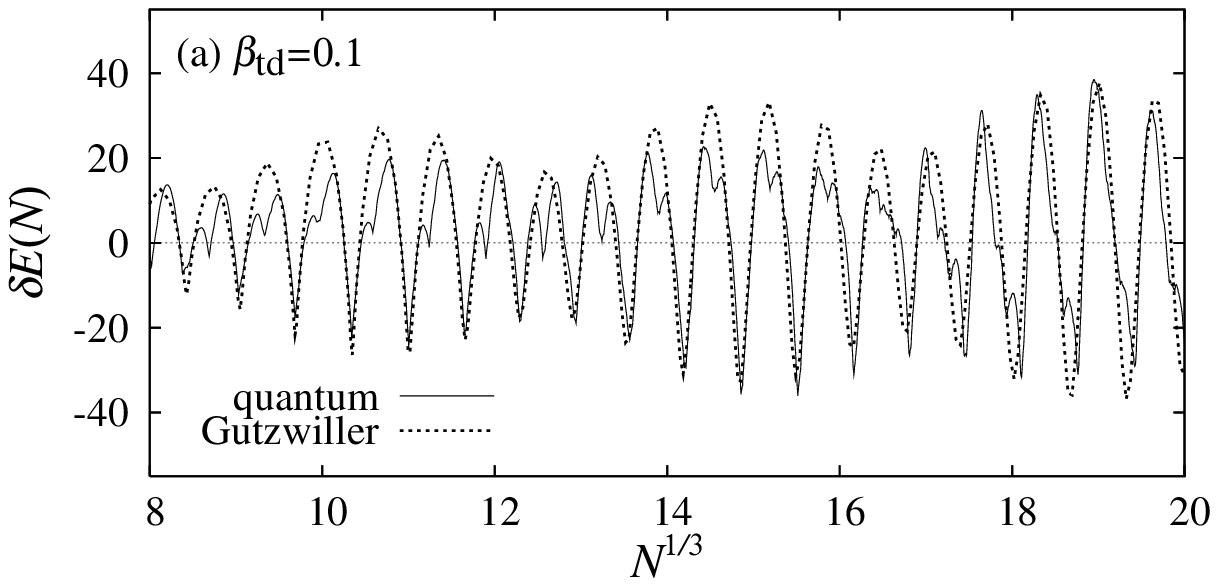}\\
\includegraphics[width=\linewidth]{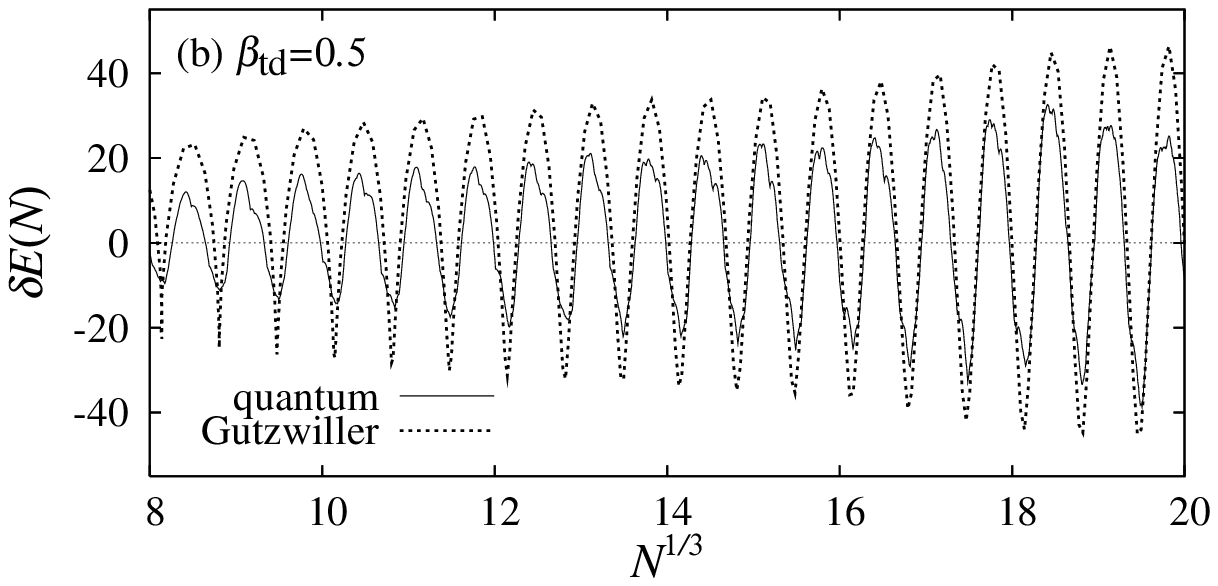}\\
\includegraphics[width=\linewidth]{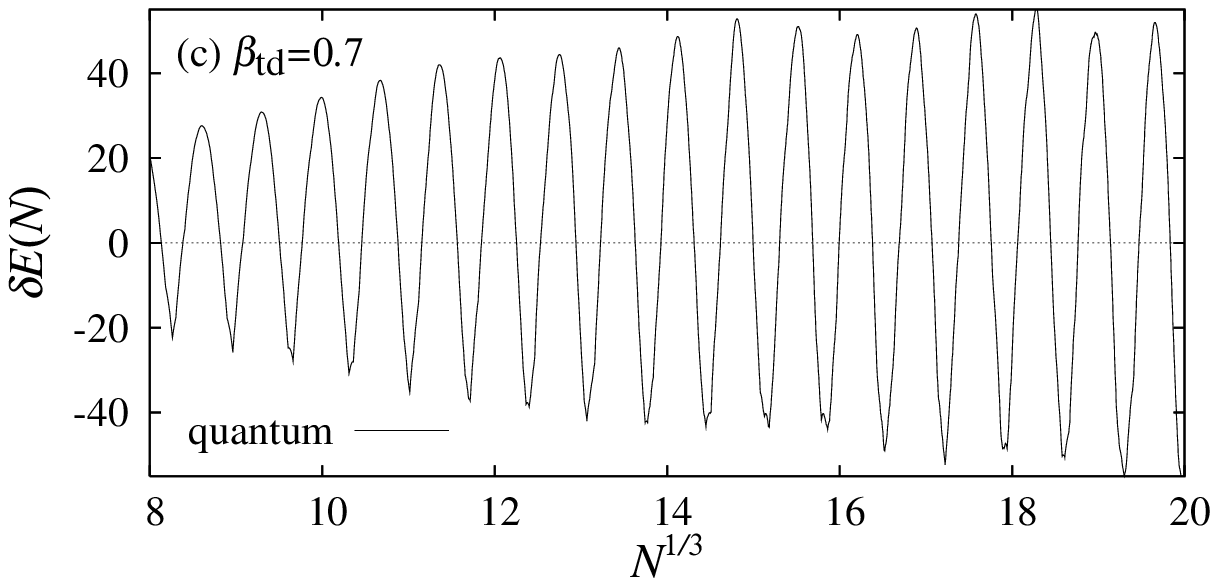}\\
\caption{\label{fig:sce}
Shell correction energies plotted as functions of the cubic root
of particle number $N$ for tetrahedral deformation parameters
$\btetra=0.1$ (a), 0.5 (b) and 0.7 (c), with radial parameter
$\alpha=6.0$.
The solid curves represent quantum results, and the dotted curves
represent semiclassical results from the Gutzwiller trace formula
with shortest periodic orbits shown in Fig.~\ref{fig:orbits}.}
\end{figure}

Figure~\ref{fig:sce} shows shell correction energies plotted as
functions of the cubic root of particle number $N^{1/3}$ for $\alpha=6.0$
with several values of $\btetra$.  Solid curves represent quantum
mechanical results, and dashed curves represent the semiclassical
results from Gutzwiller trace formula.  For $\btetra=0.1$, quantum
shell correction energies show an obvious beating pattern, which is
referred to as the supershell structure.  In trace formula, the contribution
of six orbits (DA, DB, PA, PB, TA, and TB) are taken into account.  The
above supershell structure is nicely reproduced as the result of
interference between the contribution of shorter $\tau\simeq 5.2$ and
longer $\tau\simeq 6.1$ periodic orbits.  For $\btetra=0.5$, we take
account of the contribution of new orbit TC in addition to the above six
orbits.  Here, the shorter orbits PA and DB become strongly unstable
and they do not have much contributions in the periodic orbit sum.
Thus, since the dominating four orbits have similar periods here, the
resulting shell structure shows somewhat simple oscillations.  This
also nicely reproduces the quantum result, where the supershell structure
has disappeared.  For $\btetra=0.7$, The quantum shell energies show
a regular and very strong shell effect.  Here, the Gutzwiller formula breaks
down due to the occurrence of many bifurcations, but the remarkable
enhancement of the shell effect here should be originated from coherent
contributions of many bridge orbits having
approximately the same periods.

In order to confirm the above scenarios, let us consider the Fourier
transform of the scaled-energy level density
\begin{equation}
F(\tau)=\int e^{i\cE\tau}g(\cE)e^{-\frac12(\cE/\cE_c)^2}d\cE.
\label{eq:fourier}
\end{equation}
The Gaussian with cut-off energy $\cE_c$ is included in the integrand
to suppress the contribution of level density at high energy
($\cE\gg\cE_c$) which is numerically unavailable.  This function can
be easily evaluated with the quantum mechanically obtained energy
spectra $\{\cE_i\}$ as
\begin{equation}
F^{\rm(qm)}(\tau)=\sum_i e^{i\cE_i\tau-\frac12(\cE_i/\cE_c)^2}.
\end{equation}
On the other hand, a substitution of semiclassical formula
(\ref{eq:scaled_dos}) into Eq.~(\ref{eq:fourier}) gives
\begin{equation}
F^{\rm(cl)}(\tau)\simeq F_0(\tau)+\pi\sum_{nk}
e^{-i\frac{\pi}{2}\sigma_{nk}}a_{nk}
\delta_\varDelta(\tau-n\tau_k),
\end{equation}
where $\delta_\varDelta(x)$ represents the normalized Gaussian with
the width $\varDelta=1/\cE_c$
\[
\delta_\varDelta(x)
=\frac{1}{\sqrt{2\pi}\varDelta}e^{-\frac12(x/\varDelta)^2}.
\]
Thus, the Fourier transform of level
density should exhibit peaks at the scaled periods $\tau=n\tau_k$ of
classical periodic orbits $k$ (and their repetitions) with the peak
heights proportional to the amplitude factor $a_{nk}$ of the
corresponding orbits.
\begin{figure}[tb]
\includegraphics[width=\linewidth]{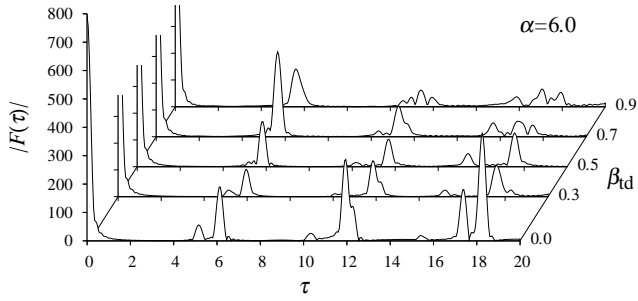}
\caption{\label{fig:fpeak}
Moduli of Fourier transform of quantum level density
for $\alpha=6.0$ with several values of $\btetra$.}
\end{figure}

Figure~\ref{fig:fpeak} shows the Fourier amplitude of the quantum level
density for $\alpha=6.0$ with several values of $\btetra$, plotted as
functions of $\tau$.  At spherical shape ($\btetra=0$), one finds peaks
at $\tau\simeq 5.1$ and $6.1$, which correspond to linear and circular
periodic orbits, respectively.  At small $\btetra$, these two
contributions interfere and build the supershell structure as shown in
Fig.~\ref{fig:sce}(a).  One also finds big peaks at larger $\tau$; for
example, the peak at $\tau\simeq 12$ corresponds to five-star orbit
bifurcated from the second repetition of the circular orbit.  These
longer orbits contribute to the finer shell structures and do not
affect the shell correction energy much.  Therefore, let us focus
on the two peaks of small $\tau$.  In increasing tetrahedral parameter
$\btetra$, the left peak corresponding to the diametric orbits rapidly
decreases, while the right peak significantly increases and take
maximum value around $\btetra=0.7$.  It is now clear from this peak
structure that the supershell structure disappears and turns into
regular oscillations as $\btetra$ increases, and the shell effect is
strongly enhanced around $\btetra\approx 0.7$ where periods of many
periodic orbits gather into approximately the same value.  This clearly
explains the behavior of the single-particle level structure in
Fig.~\ref{fig:spectrum} and of the shell correction energies in
Fig.~\ref{fig:sce}.

\begin{figure}[bt]
\includegraphics[width=\linewidth]{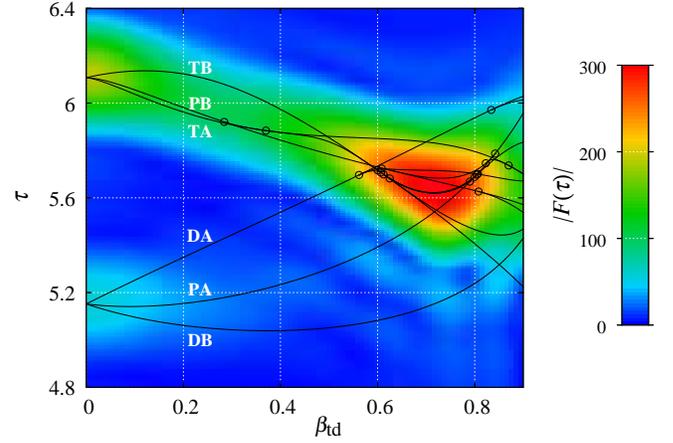}
\caption{\label{fig:ftau}
(Color online) Quantum-classical correspondence in Fourier spectra.
Quantum Fourier amplitudes $|F(\tau;\btetra)|$ are shown by the color.
Solid curves represent the scaled periods of classical periodic orbits
as functions of $\btetra$, and solid circles indicate bifurcation
points.}
\end{figure}

Figure~\ref{fig:ftau} shows the correspondence between the Fourier
transform of the quantum level density and classical periodic orbits.
One will see nice agreements between the peaks of Fourier amplitudes
and classical periodic orbits.  One should also notice that the
Fourier peak is strongly enhanced in the bifurcation region
$\btetra=0.6\sim 0.8$.  Thus we can conclude that the anomalous shell
effect emerging at large tetrahedral deformation originates from the
quasidegeneracies of the periods of several periodic orbits and the
multiple bridge bifurcations among them, which may be related to the
restoration of dynamical symmetry as discussed in previous sections.

\section{Summary and discussions}
\label{sec:summary}

Novel shell structures developed by the tetrahedral deformation have
been investigated.  We have shown that this anomalously strong
enhancement of shell effects in the nonintegral system has a close
relation to the bridge bifurcations of periodic orbits having almost
the same periods, which are similar to the situation in rational HO
system.  Its relation to the fact that the tetrahedral magic
numbers are exactly the same as those for spherical HO might also
be an interesting problem.

In evaluating the semiclassical level density, the Gutzwiller formula
overestimates the amplitude factor and we could not reproduce the shell
structures in the bifurcation region, which we are most interested in.
One of the ways to overcome this problem will be the use of uniform
approximations \cite{SS97}.  One may suspect that the bifurcations in
three-dimensional systems might be quite complicated in comparison to
the two-dimensional case where all the basic bifurcation scenarios are
classified by catastrophe theory \cite{Ozorio}.  Fortunately, we can
expect that most of the bifurcations might be analyzed in the same way
as in two-dimensional systems.  Since the monodromy matrix has a unit
eigenvalue at the bifurcation point, four eigenvalues of the monodromy
matrix of bifurcating orbits must be decomposed into two
conjugate/reciprocal pairs.  Thus, we can reduce the dimension by
decomposing the phase space into those spanned by eigenvectors
belonging to each pair of eigenvalues.  A practical procedure might be
complicated, but it is of great interest to us to estimate
semiclassical level densities and shell correction energies in the
bifurcation regions and reproduce the anomalous shell effect obtained
for the tetrahedral state.

For more quantitative discussions on nuclear systems, one should take
account of the spin-orbit coupling.  It is also an important subject
to examine if the strong shell effect obtained in this work might
survive after introducing the appropriate spin-orbit coupling, and how the
resulting shell structures might be explained in periodic orbit
theory.

\acknowledgments
We thank Matthias Brack, Alexander G.~Magner, Naoki Tajima and Kenichi
Matsuyanagi for discussions and comments.  A part of the numerical
calculations were carried out at the Yukawa Institute Computer
Facility.

\appendix
\section{Tetrahedral shape parametrization}
\label{app:shape}

Let us consider the tetrahedron whose four vertices are located at
$(-R_0,-R_0,-R_0)$, $(-R_0,R_0,R_0)$, $(R_0,-R_0,R_0)$ and
$(R_0,R_0,-R_0)$.  The equations of four faces are
\begin{equation}
\begin{split}
x+y+z=R_0, \quad x-y-z=R_0,\\
y-z-x=R_0, \quad z-x-y=R_0.
\end{split}
\end{equation}
On the surface of the tetrahedron, one of the above four equations
is satisfied, namely,
\begin{align}
(R_0-x-y-z)(R_0-x+y+z)(R_0-y+z+x)\nonumber \\
\times (R_0-z+x+y)=0.
\end{align}
It is transformed into spherical coordinates $(r,\theta,\varphi)$ as
\begin{equation}
R_0^4-2R_0^2r^2-R_0u_3(\theta,\varphi)r^3+u_4(\theta,\varphi)r^4=0,
\label{eq:shape_tetra}
\end{equation}
with
\begin{gather}
u_3(\theta,\varphi)=\frac{4}{15}P_{32}(\cos\theta)\sin 2\varphi,\\
u_4(\theta,\varphi)=\frac15+\frac45 P_4(\cos\theta)
 +\frac{1}{210}P_{44}(\cos\theta)\cos 4\varphi,
\end{gather}
where $P_{l}(x)$ and $P_{lm}(x)$ are the Legendre polynomial and
associated Legendre function, respectively.
The functions $u_3$ and $u_4$ are both symmetric under the
transformations of the tetrahedral group.  The equation of tetrahedral
surface can be obtained as the least positive root of the above
fourth-order equation of $r$.
Writing $r=R_0f(\theta,\varphi)$, this equation becomes
\begin{equation}
1-2f^2-u_3f^3+u_4f^4=0.
\label{eq:shape_tetra2}
\end{equation}
If one introduces a parameter $\btetra$ and modifies the above equation
as
\begin{equation}
f^2+\frac{\btetra}{2}(1+u_3f^3-u_4f^4)=1,
\end{equation}
$\btetra=0$ gives the sphere $(f=1)$, and $\btetra=1$ gives the equation
identical to Eq.~(\ref{eq:shape_tetra2}), namely, the tetrahedron.  Thus,
by varying the parameter $\btetra$ from 0 to 1, one can smoothly
interpolate the sphere and tetrahedron keeping the tetrahedral
symmetry.

\section{Irreps of tetrahedral group and the basis decomposition
procedure}
\label{app:group}

The tetrahedral group $\rTd$ contains 24 symmetry transformations
which are classified into five classes: identity $E$, eight rotations
$C_3$, six rotatory reflections $S_4$, three rotations $C_2$ and six
reflections $\sigma_d$.  Here we follow the notations in
Ref.~\cite{LLQMText}.  This group has five irreducible representations
(irreps): two one-dimensional irreps $\rA_1$, $\rA_2$, one two-dimensional
irrep $\rE$, and two three-dimensional irreps $\rF_2$, $\rF_1$.  The
quantum spectra can be obtained by a diagonalization of the Hamiltonian
within the bases of each irrep.  After the complete decomposition of
the bases, the dimension of the Hamiltonian matrix for irreps
$\rA$, $\rE$, and $\rF$ are about 1/24, 1/12, and 1/8, respectively, of
the total number of bases.  Thus one can highly reduce the numerical
loads by the basis decomposition.

For the construction of the basis, the irreps of the rotational group are
employed.  The basis function of the rotational group is given by
spherical harmonics $Y_{lm}$.  They form reducible representations of
the group $\rTd$, and the decomposition to each irrep can be carried
out by the projection method \cite{GroupText}.  The projection
operator onto the irrep $\alpha$ is given by
\begin{equation}
P^{(\alpha)}=\frac{f_\alpha}{g}\sum_G\chi^{(\alpha)*}(G)G,
\end{equation}
where the sum is taken over all the symmetry transformations $G$ of the
group $\rTd$, $g=24$ is the order of the group $\rTd$, $f_\alpha$ is
the dimension of the irrep $\alpha$, and $\chi^{(\alpha)}(G)$ is the
character of $G$ in the irrep $\alpha$.  By applying the
$P^{(\alpha)}$ on $Y_{lm}$, a linear combination of the bases of irrep
$\alpha$ included in $Y_{lm}$ is extracted.  $GY_{lm}$ can be
calculated by the Euler angle representation of the transformation $G$
as
\begin{equation}
R(\varOmega)Y_{lm}
=\sum_{m'}D^{(l)}_{m'm}(\varOmega)Y_{lm'},
\end{equation}
where $R(\varOmega)$ is the rotation with Euler angles $\varOmega$,
and $D^{(l)}_{m'm}(\varOmega)$ is the Wigner's $D$ function.  The
complete set of the bases in irrep $\alpha$ are obtained by extracting
all the linearly independent functions out of $2l+1$ functions
$P^{(\alpha)}Y_{lm}$ $(-l\leq m\leq l)$ by the Gramm-Schmidt
orthogonalization process.

The spectrum in irrep $\rE$ is doubly degenerated and the bases can be
further decomposed into two parts by the parity with respect to a
reflection $\sigma_d$.  The spectrum in irreps $\rF$ are triply
degenerated and the bases are also decomposed by $\sigma_d$ into two
parts: one is doubly degenerated and the other is of no degeneracies.
In obtaining spectra, one has only to consider the non-degenerated
part, which corresponds to $\sigma_d=-1$ for the
irrep $\rF_2$ and $\sigma_d=1$ for the irrep $\rF_1$.

In the current numerical calculations, we diagonalized the Hamiltonian
using the spherical harmonic oscillator bases with 41 major shells
($0\leq N_{\rm sh}\leq 40$).  The largest decomposed matrix (for the
irrep $\rF_2$) has the size 1650 among 12341 single-particle bases in
total.

\section{A simple way of calculating Maslov indices for isolated
periodic orbits in three dimensions}
\label{app:maslov}

The Maslov index $\sigma_{po}$ for a periodic orbit (PO) is practically
obtained as the sum of two contributions \cite{Gutzwiller,BBText}.
One is from the number of conjugate points $\mu_{po}$, singularities of the
semiclassical propagator, and the other is the number of negative
eigenvalues $\nu_{po}$ of the matrix
\begin{gather}
\left(\pp{^2S(\br'',\br')}{\br''\partial\br''}
+\pp{^2S(\br'',\br')}{\br''\partial\br'}
+\pp{^2S(\br'',\br')}{\br'\partial\br''}
+\pp{^2S(\br'',\br')}{\br'\partial\br'}\right)_{po},\nonumber\\
\label{eq:sqq}
\end{gather}
which arises when one evaluate the trace integral by means of a
stationary phase approximation.  Here, $S$ is the action integral
along the PO
\begin{gather}
S(\br'',\br')=\int_{\br'}^{\br''}\bp\cdot d\br,
\end{gather}
and $\br''=\br'$ is an arbitrary point on the orbit.  Each of those
two contributions ($\mu_{po}$ and $\nu_{po}$) depends on the choice of
the initial point $\br'$ or the choice of coordinate set, but the sum
$\sigma_{po}=\mu_{po}+\nu_{po}$ does not depend on such conditions.
The matrix (\ref{eq:sqq}) is directly connected to the monodromy
matrix \cite{Gutzwiller} and there are no difficulties in obtaining
$\nu_{po}$.  Let us consider the way of counting conjugate points with
respect to initial point $\br'$.  Conjugate points are classified into
three kinds of singularities: turning points, focal points and
caustics.  At the turning point, the velocity becomes zero and the orbit
forms a cusp there.  At the focal points, a group of orbits ejected from
initial point $\br'$ with momentum directions slightly deviated from
$\bp'$ concentrates into the original orbit with respect to one
direction.  At caustics, the orbit contacts with the envelope formed by
the above group of orbits.  The focal points and caustics along the
orbits correspond to the singularities of the determinant
\begin{equation}
\det\left(-\pp{^2S(\br,\br')}{\br\partial\br'}\right)_{2,3}
=\det\left(\pp{\bp'}{\br}\right)_{2,3},
\end{equation}
namely the zeros of the determinant
\begin{equation}
\det\left(\pp{\br}{\bp'}\right)_{2,3}\equiv \det\,(D)_{2,3}.
\end{equation}
The suffixes imply the local Cartesian coordinates where the first axis is
taken along the PO and the second and third axes are perpendicular to it.
Let us consider two orbits in vicinity of the PO, whose initial momentum
$\bp'$ is infinitesimally shifted towards the second and third axes,
respectively, by
\begin{equation}
\delta\bp'_2=\Matrix{0\\ \eta_2\\ 0},\quad
\delta\bp'_3=\Matrix{0\\ 0\\ \eta_3}.
\end{equation}
Then, the deviations of those orbits from PO are
\begin{gather}
\delta\br_2=\left(\pp{\br}{\bp'}\right)\delta\bp'_2
=\Matrix{D_{12}\\ D_{22}\\ D_{32}}\eta_2, \\
\delta\br_3=\left(\pp{\br}{\bp'}\right)\delta\bp'_3
=\Matrix{D_{13}\\ D_{23}\\ D_{33}}\eta_3,
\end{gather}
and one has
\begin{equation}
(\det D)_{2,3}=D_{22}D_{33}-D_{23}D_{32}
=\frac{1}{\eta_2\eta_3}(\delta\br_2\times\delta\br_3)_1.
\end{equation}
Together with turning points $|\dot{\br}|=0$, the total number of
conjugate points is obtained by counting the zeros of the quantity
\begin{equation}
U(t)=\frac{1}{|\delta\bp_2'||\delta\bp_3'|}
(\delta\br_2(t)\times\delta\br_3(t))\cdot\dot{\br}(t)
\label{eq:cjpt}
\end{equation}
along PO $(0<t<T_{po})$.
\begin{figure}[t]
\begin{center}
\includegraphics[width=\linewidth]{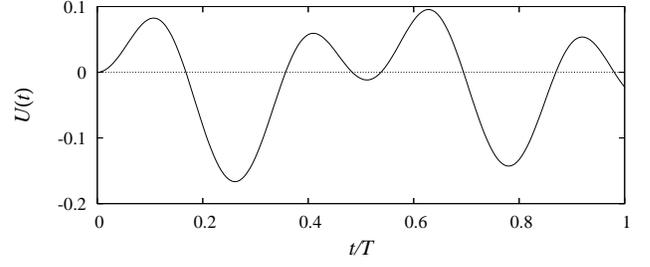}
\end{center}
\caption{\label{fig:maslov}
The function $U(t)$ of Eq.~(\ref{eq:cjpt}), calculated
for one of the orbit TB with $\alpha=6.0$ and $\btetra=0.1$.}
\end{figure}
Figure~\ref{fig:maslov} shows an example of
$U(t)$ for three-dimensional orbit TB with $\alpha=6.0$ and $\btetra=0.1$.
One obtains $\mu=7$ from this plot.

\end{document}